**Spin as the Basis for Quantum Mechanics:**

**A New Semiclassical Model for Electron Spin**

Alan M. Kadin

Princeton Junction, New Jersey, USA

April 8, 2005

**Abstract:**

A simple real-space model for the free-electron wavefunction with spin is proposed, based on coherent vortices on the scale of $\hbar/mc$, rotating at $\omega = mc^2/\hbar$. This reproduces the proper values for electron spin and magnetic moment. Transformation to a moving reference frame turns this into a wave with the de Broglie wavelength. The mapping of the real two-dimensional vector phasor to the complex plane satisfies the Schrödinger equation. This suggests a fundamental role for spin in quantum mechanics.

*E-mail amkadin@alumni.princeton.edu

## I.  Introduction

Since its inception almost a century ago, quantum mechanics has generated more than its share of mystery.  Although its precise applicability was clearly established early on, fundamental issues related to interpretation have continued to be extensively discussed. The present paper deals with the electron, and more specifically with two key properties: its complex wavefunction, and its intrinsic spin.  In the standard interpretation [1], there is no clear real-space picture of what is oscillating in the wave, or what is rotating in the spin.  Indeed, it is generally believed that no simple model of rotation can account for the spin of the electron.  On the contrary, the present paper shows that a crude mechanical model of coherently rotating vortices can account quantitatively not only for spin, but also for the wavefunction itself (a preliminary version of this model is available online in [2]).  The mathematics of a rotating vector field are equivalent to those of a complex scalar field, suggesting that this rotating spin field *is* the quantum mechanical wavefunction.  This further suggests that spin is central to quantum mechanics, rather than being a separate feature that is present only in certain cases.  The implications of this are discussed later in the paper.

## II.  The Model

First, consider an electron with its center of mass at rest, but spinning.  The simplest possible model (which will be modified in the next paragraph) is a spinning solid sphere. Based on the goal of having this describe the electron wavefunction, one expects that the angular velocity is given by the Planck-Einstein relation $E = \hbar\omega$.  Since this is a real physical rotation, the zero of energy is not arbitrary as in standard nonrelativistic quantum mechanics, but must be given by the relativistic rest energy $E=mc^2$.  (This also has the property of being relativistically covariant when we transform later to a moving reference frame.)  For rotation of a solid sphere of radius R, the linear velocity on the equator is $u=R\omega = Rmc^2/\hbar$.  But clearly, u can be no greater than the speed of light c. This is a natural cutoff, and provides an estimate of the maximum size of this spinning ball:

$$R_{max} = c/\omega = \hbar/mc = R_c. \qquad (1)$$

This is the Compton wavelength $R_c$ of the electron $\sim 0.4$ pm, which is much smaller than the typical Å scale that characterizes atomic orbitals (1 Å = 100 pm). If we want to model an extended electron state, then clearly $R_c$ is too small.

Consider instead an extended state consisting of a parallel array of cylindrical vortices (see Fig. 1), each a solid body of radius $R_c$ rotating around its axis at $\omega = mc^2/\hbar$. For simplicity here, assume that there are N identical vortices, each of mass $m_v = m/N$. The angular momentum of each vortex is then given by

$$L_v = I\omega = \tfrac{1}{2} m_v R_c^2 \,\omega = \hbar/2N, \qquad (2)$$

where we have taken the moment of inertia $I = \tfrac{1}{2} mR^2$ for a cylinder of uniform mass density. This is a crude semi-relativistic model, but it does in fact give the proper value for the total angular momentum for the electron, $S = \hbar/2$.

One can also estimate the magnetic moment of the electron from this model. Treating the rotating charge per vortex $q_v = e/N$ as a current $i_v = q_v\omega/2\pi$, one obtains simply

$$\mu = Ni_vA_v = (e\omega/2\pi)(\pi R_c^2) = e\hbar/2m = \mu_B, \qquad (3)$$

where $\mu_B$ is the Bohr magneton and $A_v$ is the cylindrical cross sectional area per vortex. Again, this is the correct result, perhaps fortuitously, but it does suggest that this crude model may incorporate much of the essential physics.

These calculations require only that all of the cylinders are rotating at the same frequency around parallel axes. But in addition, it is reasonable to assume a coherent state where all of them are rotating in-phase as well, as suggested in Fig. 1. This requires a rotating vector field $\mathbf{F}(\mathbf{r},t)$. Furthermore, it is not necessary to assume that the vortices have identical masses. More generally, one could have a density function $\rho(\mathbf{r})$, which would go as the square of the field amplitude $F(\mathbf{r})$, analogously to the energy density in electromagnetic waves.

Now the phase angle θ(t) = Et/ℏ is constant across the entire electron, but that can also be relaxed. Consider what happens when we Lorentz-transform to a reference frame moving with velocity **v**. Locations that are in phase in the rest frame will not in general be in phase in the moving frame. The proper way to deal with this is to make the phase angle relativistically invariant, so that

$$Et \Rightarrow E't' - \mathbf{p'} \cdot \mathbf{r'}, \tag{4}$$

where in the usual way $E' = \gamma mc^2 \approx mc^2 + \frac{1}{2} mv^2$, $\mathbf{p'} = \gamma m\mathbf{v} \approx m\mathbf{v}$ is the momentum, $\gamma = (1 - v^2/c^2)^{-\frac{1}{2}}$, and the approximate forms are for v<<c. This is invariant because $(E/c, \mathbf{p})$ and $(ct, \mathbf{r})$ are relativistic 4-vectors, and the phase angle goes as their inner product. So now the rotating phase angle takes the form

$$\theta(\mathbf{r}, t) = (Et - \mathbf{p} \cdot \mathbf{r})/\hbar \tag{5}$$

This corresponds to a plane wave with wavelength λ = h/p, which is well known as the de Broglie wavelength. Note that this follows directly from the earlier assumption that the rotation frequency is given by $mc^2/h$.

Once we have a wave satisfying the Einstein-deBroglie relations, the rest of quantum mechanics follows naturally. We have a rotating vector field given by a spin axis (assumed to be uniform), an amplitude F(**r**,t), and a rotating phase angle θ(**r**,t). If we compare to the standard complex wavefunction in quantum mechanics, Ψ(**r**,t) = |Ψ|exp(iφ), and map F and θ onto |Ψ| and φ, we have a rotating wavefunction which satisfies the time-dependent Schrödinger equation.

For example, consider a rotating vector field of the form

$$\mathbf{F}(r, t) = F[\mathbf{u_x} \cos(kz - \omega t) \pm \mathbf{u_y} \sin(kz - \omega t)], \tag{6}$$

(**u_x** and **u_y** are the unit vectors in the x- and y-directions), which represents a plane wave traveling in the z-direction with spin also in the z-direction. This is a circularly polarized transverse wave, with either positive or negative helicity depending on whether the plus or minus sign is chosen. For fixed t, the tip of the vector follows a helix; for fixed z, circular rotation at an angular frequency $\omega$ of a vector of length F. Now define $\theta = \arctan(F_y/F_x) = kz-\omega t$, and

$$\Psi(r,t) = F\exp(i\theta) = F\exp[i(kz-\omega t)], \tag{7}$$

and substitute this into the time-dependent Schrödinger equation with the rest-energy explicitly added:

$$i\hbar\ \partial\Psi/\partial t = H\Psi = (-\hbar^2/2m)\nabla^2\Psi + [mc^2+V(r)]\ \Psi. \tag{8}$$

The result is the simple, correct relation (for v<<c) that $\hbar\omega = \hbar^2k^2/2m + mc^2$. Note also that the complex conjugate of $\Psi$ might seem to yield negative energy, but really just represents the spin of the opposite sign.

Thus far the model has been limited to a single plane wave, but electrons are generally present in bound states, with standing waves instead of traveling waves. Consider for simplicity the one-dimensional particle-in-a-box, with the electron confined between z=0 and z=L. The solution takes the form of discrete bound states given by the complex wavefunctions $\Psi_n$ and equivalent vector fields $\mathbf{F}_n$:

$$\Psi_\mathbf{n} = \sin(n\pi z/L)\exp(-i\omega t) \tag{9}$$

$$\mathbf{F}_\mathbf{n} = \sin(n\pi z/L)(\mathbf{u}_\mathbf{x}\ \cos\omega t \pm \mathbf{u}_\mathbf{y}\ \sin\omega t). \tag{10}$$

Here n=1 corresponds to the ground state and n=2, 3,... to the excited states, and the quantized energies $E_n$ are given as usual (but with the $mc^2$ offset) by

$$E_n = \hbar\omega_n = mc^2 + \hbar^2k^2/2m = mc^2 + \hbar^2(n\pi/L)^2/2m \tag{11}$$

and as before the ± corresponds to the two spin states. Note that this vector wavefunction has separated into two factors, the usual standing-wave envelope and the rotating phase vector. The negative values of the sine (for n>1) correspond to 180º shifts of the rotating phase.

### III. Discussion and Conclusions

The wave example given above is based on a helical transverse wave, which is similar in form to a transverse electromagnetic wave which is circularly polarized. Indeed, such a helical TEM wave carries angular momentum, and forms the classical limit of a photon [3,4], with spin $\pm\hbar$ pointing along the direction of motion. However, unlike the case of the photon, one can transform to the rest frame of the electron, and from there to any other direction. In general, the electron spin axis would not be parallel to the momentum, and the rotating spin field vector would follow a general cycloidal motion rather than a simple helix. The spin and translational motions are essentially decoupled in this model (no spin-orbit interaction).

This model of coherently rotating vortices appears to account for the complex wavefunction of the electron [2]. This suggests that the spin picture may be substantially more general than simply a single electron, and that spin is fundamental to all of quantum mechanics. In that regard, it may not be a coincidence that all fundamental quantum particles seem to have spin. Certain mesons have spin-0, but they can be regarded as composites of spin-½ quarks. And certainly atoms with spin-0 show quantum effects. It is likely that the spins of the constituent components contribute their angular phase references to the composite system, even if the total spin cancels out.

One may speculate as to the physical basis for such a coherent vortex model. It seems to correspond to a very rigid state of an intrinsically rotating fluid. Such a rigid state may indicate a very strong cohesive energy associated with long-range phase coherence among the vortices. Since the lowest excitation of an electron involves creation of an electron-positron pair, this cohesive energy might be expected to be ~ 1 MeV, larger than the rest energy of the electron itself.

Speculating even further, the existence of such a highly rigid state would have important implications for quantum measurement. Any local interaction that would alter the energy of part of an electron wavefunction would jeopardize this cohesive energy. This, in turn, would create an instability leading either to the rest of the electron being pulled into the interaction region, or alternatively to the expulsion of the electron from this region. This suggests a real dynamical process which may provide a physical basis for the "collapse of the wavefunction" in quantum measurement.

Finally, if this rotating spin field is mathematically equivalent to the usual Schrödinger equation, is it really just a matter of preference which representation we choose? Not entirely, because a real physical rotation, with a definite frequency and spatial fine structure, should be measurable. If one probes the behavior of electrons at frequencies ~ $10^{20}$ Hz = $mc^2/h$, particularly with a circularly polarized probe, one should expect to see a sharp resonance in some sort of spectral response, perhaps associated with spin-flip of the electron in a large magnetic field. Furthermore, the fine structure of the spin model identified a periodicity on the scale of $2R_c = 2\hbar/mc$, which would correspond to a momentum transfer $\hbar k = \pi mc \sim 1.5$ MeV/c. It would be interesting to see whether relevant measurements are consistent with the model described in this paper.

It is somewhat surprising that a simple mechanical model for spin was not presented in the early days of quantum mechanics. It seems that early researchers were discouraged by apparent rotation velocities greater than c [3]. It may be that the distributed coherent vortex model provides a way around these difficulties. More recently, Ohanian [3] showed that the relativistic Dirac equation is consistent with a distributed circular energy flow on a scale larger than $R_c$, which provides the basis for the electron spin and magnetic moment. The present semiclassical model is certainly cruder than the Dirac equation, but also reproduces these results within a more intuitive physical picture.

In conclusion, a new semiclassical picture for electron spin is presented, in which a spinning vector field, rotating at $mc^2/h$, is organized into a coherent array of rigidly rotating vortices on the scale of $R_c = \hbar/mc$. The vector field **F** maps onto the quantum wavefunction $\Psi$, providing for a unification of spin and quantum mechanics. It is further suggested that the coherent nature of this spin field may be associated with a cohesive energy, which in turn may play a key role in quantum measurement. While the specific details of this model remain crude, its clear intuitive physical picture may help to stimulate further research along similar lines. By dealing with specific real-space models, it may be possible to remove much of the abstraction and mystery from quantum theory.

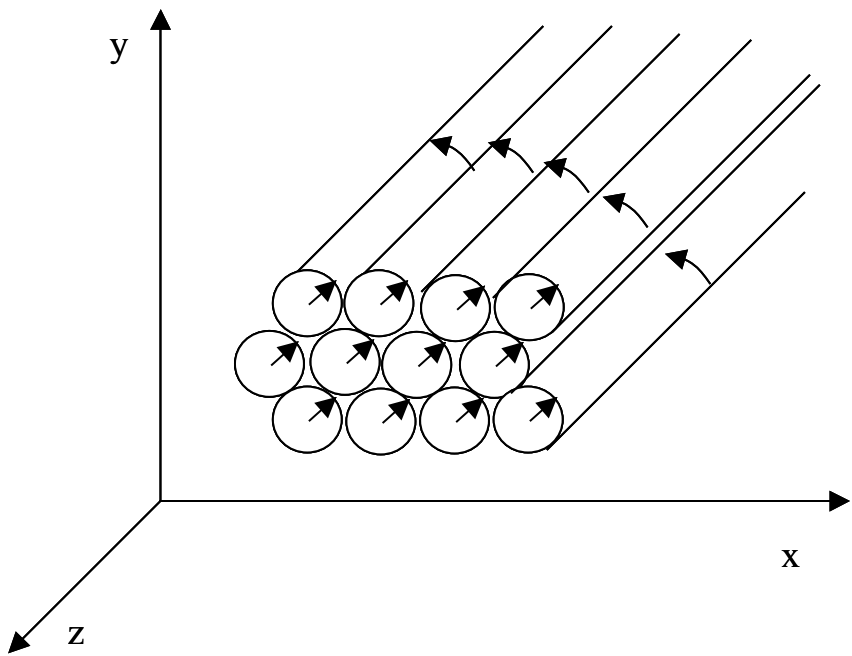

Fig. 1. Picture of coherent vortex model of electron spin, with fine structure of parallel array of vortices on scale of $R_c = \hbar/mc = 0.4$ pm. Here the spin is pointing in the z-direction.